\documentclass[final, runningheads]{llncs}
\usepackage{url}
\usepackage[T1]{fontenc}

\usepackage{listings}
\usepackage{color}
\definecolor{mygreen}{rgb}{0,0.6,0}
\definecolor{mygray}{rgb}{0.5,0.5,0.5}
\definecolor{mymauve}{rgb}{0.58,0,0.82}
\definecolor{code_background}{rgb}{1,0.94,0.8}

\newcommand{\synergy}[1]{\textsc{SYnergy}} 
\newcommand{\sbench}[1]{\text{23}}

\usepackage{orcidlink} 

\usepackage{graphicx}
\usepackage{amsmath}
\newcommand{\itadata}{\footnotesize \textsl{Workshop Scientific HPC in the pre-Exascale era (part of ITADATA2024)}}
\usepackage{fancyhdr}
\fancypagestyle{empty}{\fancyhf{}\fancyhead[C]{\itadata}
  \fancyhead[R]{\includegraphics[width=.05\textwidth]{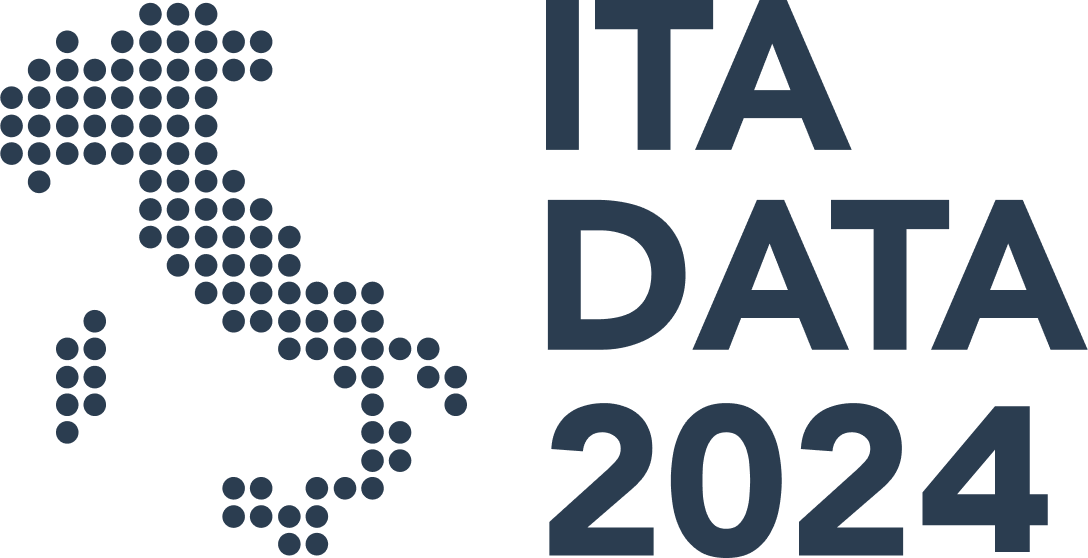}}}

\makeatletter
\begin{document}
\title{Toward Heterogeneous, Distributed, and Energy-Efficient Computing with SYCL}

\author{
Biagio Cosenza\inst{1}\orcidlink{0000-0002-8869-6705} \and
Lorenzo Carpentieri\inst{1}\orcidlink{0009-0001-2041-7618} \and
Kaijie Fan\inst{1}\orcidlink{0000-0003-0118-0974} \and
Marco D'Antonio \inst{2}\orcidlink{0009-0002-7399-7705} \and
Peter Thoman \inst{3}\orcidlink{0000-0002-4028-7451} \and
Philip Salzmann \inst{3}\orcidlink{0000-0002-8668-4639}
}
\authorrunning{Cosenza et al.}
\institute{University of Salerno, Italy \and
Queen's University Belfast, UK \and
University of Innsbruck, Austria}

\maketitle              \begin{abstract}
Programming modern high-performance computing systems is challenging due to the need to efficiently program GPUs and accelerators and to handle data movement between nodes. The C++ language has been continuously enhanced in recent years with features that greatly increase productivity. In particular, the C++-based SYCL standard provides a powerful programming model for heterogeneous systems that can target a wide range of devices, including multicore CPUs, GPUs, FPGAs, and accelerators, while providing high-level abstractions.
This presentation introduces our research efforts to design a SYCL-based high-level programming interface that provides advanced techniques such as task distribution and energy optimization. The key insight is that SYCL semantics can be easily extended to provide advanced features for easy integration into existing SYCL programs. In particular, we will highlight two SYCL extensions that are designed to deal with workload distribution on accelerator clusters (Celerity) and with energy-efficient computing (SYnergy).

\end{abstract}
\section{Introduction}
Programming models bridge the gap between the underlying hardware architecture and the supporting software layers available to applications. 
They typically focus on increasing developer productivity, performance, and portability to other systems.
Particularly in the context of High Performance Computing (HPC), the rapidly changing nature of computing architectures and the evolving complexity of the next generation of exascale computing systems pose significant challenges to these goals. 
Programming models for HPC must address very specific issues such as representing and managing increasing levels of parallelism, concurrency, and memory hierarchies; the heterogeneous nature of emerging hardware such as GPUs; and the ability to maintain increasing levels of interoperability with today's applications.
Therefore, it has become critical for HPC to have a programming environment that is capable of addressing modern heterogeneous distributed computing systems, possibly based on open standards to avoid vendor lock-in that compromises code portability.

\section{SYCL for Heterogeneous Programming}
SYCL \cite{SYCL2020} is a programming model for heterogeneous computing that builds on modern C++.  While SYCL follows the execution and memory model of OpenCL \cite{OpenCLapi,OpenCLlang}, implementations can have a non-OpenCL mapping, e.g., the AdaptiveCpp CUDA backend~\cite{hipsycl}. 
SYCL supports single-source programming where both kernel and host codes are stored in the same source file. 
In a SYCL program, the kernel code to be executed on a device such as a GPU, is expressed by the \texttt{parallel\_for} function from the command group handler object. 
The device on which the kernel code executes is represented by a queue.

Listing \ref{lst:saxpy} shows a SAXPY kernel in SYCL. A queue is created for a submitting task to the GPU. 
A kernel is essentially a lambda function provided to a \texttt{parallel\_for}.
This specific code snippet leverages on the buffer/accessor mode, where buffer are data that can be accessed from a device, e.g., a GPU, and the accessor declares to which buffer a kernel access to. 

\begin{lstlisting}[caption={SAXPY kernel in SYCL},label={lst:saxpy}]
queue q(gpu_selector);
q.submit([&] (handler& h) {
  accessor x{x_buf,h,read_only};
  accessor y{y_buf,h,read_only};
  accessor z{z_buf,h,write_only};
  h.parallel_for(range<1>(N), 
    [=](id<1> i){
      z[i] = alpha * x[i] + y[i];
    });
});
\end{lstlisting}

The most important implementations of the SYCL standard include OneAPI Data-Parallel C++ \cite{AshbaughBBHKPSS20}, developed by Intel, and AdaptiveCPP, developed at Heidelberg University, with several other minor implementations actively under development \cite{neoSYCL}. 
Nevertheless, SYCL provides a robust foundation where developers can write highly specialized code in a more accessible manner, aiming to achieve performance portability across a wide range of hardware architectures.
We have performed extensive benchmarking of SYCL 1.2.1 \cite{SYCLBench121} and 2020 \cite{SYCLBench2020} features on various platforms and compilers, showing unprecedented results in terms of stability and performance portability. 

Furthermore, the flexibility of SYCL has given rise to various extensions targeting specific heterogeneous use cases, such as Celerity for distributed computing \cite{ThomanSCF19} and SYnergy for energy optimization \cite{FanDCCFC23}.

\newpage

\section{Celerity for Distributed Computing}

Celerity \cite{ThomanSCF19,SalzmannKTGCF23} is an API and distributed runtime system designed to extend the SYCL programming model to distributed computing environments. 
The overall goal of Celerity is to bridge the gap between single-node SYCL programming and distributed cluster computing.

Celerity's API is heavily inspired by SYCL, making it familiar to developers with previous experience in SYCL programming. This design choice allows for easier adoption and migration of existing SYCL codebases. In fact, the Celerity API redefines most SYCL semantics for distributed computing, in particular through its distributed queue and a new parameter to the accessor, called the range mapper, which ties a kernel's execution range to its data requirements.
Unlike SYCL, which targets single-node systems, Celerity is built from the ground up to support distributed memory clusters. This is possible thanks to a distributed runtime system that handles the complexities of data distribution and task scheduling across multiple nodes. 
The runtime automatically handles data transfer between nodes based on the requirements expressed through the range mapper associated with each accessor.
Celerity aims to provide a higher level of abstraction for distributed memory programming, reducing the need for explicit message passing (e.g., via MPI) in many scenarios.
Internally, the runtime system uses SYCL as its underlying execution engine, allowing it to take advantage of SYCL's device-agnostic approach to accelerator programming.

\begin{lstlisting}[caption={SAXPY kernel in Celerity},label={lst:saxpycelerity}]
distr_queue q(gpu_selector);
q.submit([&] (handler& h) {
  accessor x{x_buf,h,one_to_one,read_only};
  accessor y{y_buf,h,one_to_one,read_only};
  accessor z{z_buf,h,one_to_one,write_only};
    h.parallel_for(range<1>(N), 
      [=](id<1> i){
        z[i] = alpha * x[i] + y[i];
      });
});
\end{lstlisting}

Listing \ref{lst:saxpycelerity} shows a SAXPY kernel in Celerity. A distributed queue is created for a submitting tasks to the cluster of GPUs. 
As in SYCL, a Celerity kernel is written as a lambda function passed to \texttt{parallel\_for}.
In this example, a \texttt{one\_to\_one} range mapper is provided to each accessor, indicating that the kernel will read and write only at the current index \texttt{i}. The kernel is then automatically split across all available devices, with the range mappers being used to infer the set of data transfers required before the kernel can be executed.

\section{SYnergy for Energy-efficient Computing}
Energy-efficient computing has become a crucial area of focus in High-Performance Computing (HPC) as the demand for computational power continues to rise alongside growing concerns over energy consumption and environmental impact.
Modern HPC systems require vast amounts of energy to perform large-scale computations, making power efficiency a key consideration in both hardware and software design.
One of the most important software techniques to decrease energy consumption is frequency scaling, which dynamically adjusts the clock frequency of processors in response to different applications \cite{fan2019predictable} and workloads \cite{carpentieri2023domain}.
These include the NVIDIA Management Library (NVML)\cite{NvidiaNvml}, which provides different functionalities for changing frequency and reading energy on NVIDIA GPUs; ROCm SMI \cite{RocmSmi}  , which offers similar capabilities for AMD GPUs ; and Intel’s Level Zero \cite{openapilevelzero}, designed to manage frequency and power settings on Intel GPUs.
These tools enable developers and system administrators to optimize performance while reducing energy consumption, which makes them essential for achieving sustainable and energy-efficient computing in HPC environments.
However, the vendor's power/energy interfaces are very different from each other, and there is no common interface to provide portable common functionalities. 
\synergy{} fills this gap by implementing an energy interface for coarse and fine-grained frequency scaling and energy profiling.
Inspired by the Celerity SYCL extension, \synergy{} is distributed as an SYCL header-only library, which can be easily incorporated into existing C++ build environments.
This paper uses the interface bindings for NVIDIA, AMD and Intel GPUs, which are mapped into the NVML \cite{NvidiaNvml} , ROCm SMI \cite{RocmSmi} and Level Zero \cite{openapilevelzero}, respectively.

The list \ref{synergy_code} shows a SYCL code enriched with energy profiling and frequency scaling capabilities provided by \synergy{}.
The main entry point for using the SYnergy interface is the synergy::queue class (line 1), which extends the standard SYCL queue with energy capabilities. 
The \synergy{} queue allows us to specify an energy target that defines how to set the device frequency. For example, with MIN\_EDP \synergy{} automatically selects the frequency that minimizes the Energy Delay Product (EDP).
The SYnergy API provides coarse-grained and
fine-grained energy profiling capabilities that allow measuring,
respectively, the energy consumption of the whole device (line 12) and the
energy consumption of each kernel (line 11) executed in a SYCL application.
\begin{lstlisting}[label=synergy_code,caption={A SAXPY kernel in SYnergy}]
synergy::queue q(gpu_selector, MIN_EDP);
event e = q.submit([&] (handler& h) {
accessor x{x_buf,h,read_only};
accessor y{y_buf,h,read_only};
accessor z{z_buf,h,write_only};
  h.parallel_for(range<1>(N), 
    [=](id<1> i){
      z[i] = alpha * x[i] + y[i];
    });
});
double kernel_energy = q.kernel_energy_consumption(e);
double device_energy = q.device_energy_consumption();
\end{lstlisting}

\section{Conclusions}

SYCL is a promising programming model for targeting modern heterogeneous computing systems. It is based on modern C++ and is a royalty-free standard that allows to achieve a high degree of performance portability, as demonstrated by several industrial applications.
In addition, SYCL has shown an unprecedented extensibility and flexibility, which makes it very interesting for building advanced extensions.
In fact, SYCL has been the subject of several extensions targeting specific heterogeneous use cases, such as Celerity for distributed computing and SYnergy for energy optimization.
In particular, Celerity builds on the semantics of SYCL by extending the queue and the accessor to allow a SYCL code to run on a cluster of accelerators. Similarly, SYnergy extends the SYCL queue to provide energy scaling and measurement capabilities.

In future work, we plan to further extend the language to enable new features such as support for approximate computing and AI workloads.

\begin{credits}
\subsubsection{\ackname} This work has received funding from the European High-Performance Computing Joint Undertaking under grant agreement No. 956137 (LIGATE project), 
from the Austrian Research Promotion Agency (FFG) via the UMUGUC project (FFG \#4814683) and from the Italian Ministry of University and Research under PRIN 2022 grant No. 2022CC57PY (LibreRT project), and  
from the National Recovery and Resilience Plan (NRPP), call for tender No. 104 published on 02/02/2022 by the Italian Ministry of University and Research (MUR) funded by the European Union - Next Generation EU, Mission 4, Component 1, CUP D53D23008590001, project title LibreRT.
We acknowledge the CINECA award under the ISCRA initiative (ISCRA IsCb3, project EMPI) for the availability of HPC resources and support. 
We thank Intel for hardware donation and Intel Cloud access.

\end{credits}
\bibliographystyle{splncs04}

\end{document}